\documentclass{aastex63}

\usepackage[utf8]{inputenc}  
\usepackage[T1]{fontenc}
\usepackage{graphicx,psfrag}
\usepackage{mathrsfs}
\usepackage{amsmath,amsfonts,amssymb,pifont,gensymb}
\usepackage{multirow,enumerate}
\usepackage{comment,hyperref}
\usepackage{xcolor}
\usepackage{acronym}
\usepackage{xspace}
\usepackage[normalem]{ulem}
\usepackage{mathtools}
\usepackage{subfigure}
\usepackage{makecell}

\newcommand{\red}[1]{{\textcolor{black}{#1}}}

\newcommand{\NSBHoneMmax}{2.53^{+0.07}_{-0.06} M_{\odot}} 
\newcommand{\NSBHtwoMmax}{2.68^{+0.14}_{-0.11} M_{\odot}} 
\newcommand{\BBHoneMmax}{2.18^{+0.15}_{-0.13} M_{\odot}} 
\newcommand{\BBHtwoMmax}{2.29^{+0.26}_{-0.21} M_{\odot}} 

\newcommand{\NSBHoneRonefour}{12.56^{+0.38}_{-0.51} \rm{ km}} 
\newcommand{\NSBHtwoRonefour}{12.69^{+0.55}_{-0.55} \rm{ km}} 
\newcommand{\BBHoneRonefour}{11.76^{+0.90}_{-0.82} \rm{ km}} 
\newcommand{\BBHtwoRonefour}{11.97^{+0.87}_{-0.93} \rm{ km}} 

\newcommand{\NSBHoneRonefourSim}{12.6^{+0.4}_{-0.5} \rm{ km}} 
\newcommand{\NSBHtwoRonefourSim}{12.7^{+0.6}_{-0.5} \rm{ km}} 
\newcommand{\BBHoneRonefourSim}{11.8^{+0.9}_{-0.8} \rm{ km}} 
\newcommand{\BBHtwoRonefourSim}{12.0^{+0.9}_{-0.9} \rm{ km}} 

\begin{document} 



\title{On the nature of GW190814 and its impact on the understanding of supranuclear matter}

\correspondingauthor{Ingo Tews}
\email{itews@lanl.gov}

\author[0000-0003-2656-6355]{Ingo Tews}
\affiliation{Theoretical Division, Los Alamos National Laboratory, Los Alamos, NM 87545, USA}

\author[0000-0001-7041-3239]{Peter T.~H.~Pang}
\affiliation{Nikhef, Science Park 105, 1098 XG Amsterdam, The Netherlands}
\affiliation{Department of Physics, Utrecht University, Princetonplein 1, 3584 CC Utrecht, The Netherlands}

\author[0000-0003-2374-307X]{Tim Dietrich}
\affiliation{Institut f\"{u}r Physik und Astronomie, Universit\"{a}t Potsdam, 
Haus 28, Karl-Liebknecht-Str. 24/25, D-14476, Potsdam, Germany}

\author[0000-0002-8262-2924]{Michael W.~Coughlin}
\affiliation{School of Physics and Astronomy, University of Minnesota,
Minneapolis, MN 55455, USA}

\author[0000-0002-7686-3334]{Sarah Antier}
\affiliation{Universit\'e de Paris, CNRS, Astroparticule et Cosmologie, F-75013 Paris, France}

\author[0000-0002-8255-5127]{Mattia Bulla}
\affiliation{Nordita, KTH Royal Institute of Technology and Stockholm University, 
Roslagstullsbacken 23, SE-106 91 Stockholm, Sweden}

\author{Jack Heinzel}
\affiliation{Carleton College, Northfield, MN 55057, USA}
\affiliation{Artemis, Universit\'e C\^ote d'Azur, Observatoire C\^ote d'Azur, CNRS, CS 34229, F-06304 Nice Cedex 4, France}

\author[0000-0002-5478-2170]{Lina Issa}
\affiliation{Nordita, KTH Royal Institute of Technology and Stockholm University, 
Roslagstullsbacken 23, SE-106 91 Stockholm, Sweden}
\affiliation{Universit\'e Paris-Saclay, ENS Paris-Saclay, D\'epartement de Phyisque, F-91190, Gif-sur-Yvette, France}

\date{\today}

\begin{abstract}
The observation of a compact object with a mass of $2.50-2.67M_{\odot}$ on August 14, 2019, by the LIGO Scientific and Virgo collaborations (LVC) has the potential to improve our understanding of the supranuclear equation of state. 
While the gravitational-wave analysis of the LVC suggests that GW190814 likely was a binary black hole system, the secondary component could also have been the heaviest neutron star observed to date. 
We use our previously derived nuclear-physics-multimessenger astrophysics framework to address the nature of this object. 
Based on our findings, we determine GW190814 to be a binary black hole merger with a probability of $>99.9\%$.
Even if we weaken previously employed constraints on the maximum mass of neutron stars, the probability of a binary black hole origin is still $\sim 81\%$. 
Furthermore, we study the impact that this observation has on our understanding of the nuclear equation of state by analyzing the allowed region in the mass-radius diagram of neutron stars for both a binary black hole or neutron star--black hole scenario.
We find that the unlikely scenario in which the secondary object was a neutron star requires rather stiff equations of state with a maximum speed of sound $c_s\geq \sqrt{0.6}$ times the speed of light, while the binary black hole scenario does not offer any new insight.
\end{abstract}



\section{Introduction}

Neutron stars (NSs) are the densest objects in the observable universe and allow us to probe matter under the most extreme conditions realized in nature.
In particular, NSs close to the maximum mass, i.e., the highest mass $M_{\rm{max}}$ that can be supported against gravitational collapse to a black hole (BH), truly probe matter at its limits. 
Even though NS masses could historically be inferred quite accurately through timing measurements~\citep{Lattimer:2012nd}, the exact value of $M_{\rm{max}}$ is still not known. 
For a long time, because observed NSs had masses around $1.4 M_{\odot}$, one assumed that $M_{\rm{max}}$ was not much higher. 
However, this situation changed with several observations of pulsars with $M\sim 2 M_{\odot}$ in the last decade:
PSR 1614-2230 with $M=1.908\pm 0.016 M_{\odot}$~\citep{Demorest:2010bx, Arzoumanian:2017puf}, 
PSR J0348+0432 with a mass of $M=2.01\pm 0.04 M_{\odot}$~\citep{Antoniadis:2013pzd},
and MSP J0740+6620 with a mass of $M=2.14\pm 0.10 M_{\odot}$~\citep{Cromartie:2019kug}. 
These observations firmly established that the equation of state (EOS) of NSs has to be sufficiently stiff to support such heavy stars.
Combining the likelihoods for these three observations, they provide a strong lower bound $M_{\rm{max}}\geq 2.03 M_{\odot}$ at 90\% confidence~\citep{Dietrich:2020lps}.  
An upper bound on $M_{\rm{max}}$, on the other hand, is impossible to obtain from NS mass measurements alone.
Assuming that BH and NS mass distributions do not overlap, it might be extracted from population studies or observations of BHs, e.g., \cite{Alsing:2017bbc}, \cite{Fishbach:2020ryj}, and \cite{Farr_2020}, or from nuclear-physics considerations, e.g., \cite{Kalogera:1996ci}.

In addition, the first observation of a binary neutron-star (BNS) merger, GW170817~\citep{TheLIGOScientific:2017qsa, Abbott:2018wiz} performed by the LIGO Scientific and Virgo collaborations (LVC), and the observations of the associated kilonova, AT2017gfo, and the short gamma-ray burst, GRB170817A~\citep{GBM:2017lvd} led several groups to propose upper limits on $M_{\rm{max}}$, e.g.,~\cite{Margalit:2017dij}, and \cite{Rezzolla:2017aly}.
These bounds are based on the conjecture that the ejecta properties disfavor both a prompt collapse to a BH as well as a long-lived NS.
This delayed-collapse scenario, with an expected remnant lifetime of several $100$~ms~\citep{Gill:2019bvq}, provides an upper limit on $M_{\rm{max}}$, because larger $M_{\rm{max}}$ typically lead to longer remnant lifetimes, see, e.g., \cite{Dietrich:2020eud} and references therein.
Given the observed remnant mass of $M_R = 2.7 M_{\odot}$~\citep{TheLIGOScientific:2017qsa, Abbott:2018wiz}, limits on $M_{\rm{max}}$ have been proposed in the range of $2.3-2.4~M_{\odot}$, see, e.g., \cite{Margalit:2017dij}, \cite{Ruiz:2017due}, \cite{Rezzolla:2017aly}, and \cite{Shibata:2019ctb}.
While generally robust, these upper limits on $M_{\rm{max}}$ are based on numerical simulations and on the reasonable but unproven assumption that the final remnant was a BH. 

A recent detection by Advanced LIGO~\citep{TheLIGOScientific:2014jea} and Advanced Virgo~\citep{TheVirgo:2014hva} adds a fascinating new piece of information to this puzzle. 
In its third observing run, on August 14, 2019, the LVC discovered gravitational waves (GWs) from a binary compact-object merger of a $22.2-24.3\,M_{\odot}$ BH with an unidentified compact companion of $2.50-2.67M_{\odot}$~\citep{Abbott:2020khf}.
While in the future, gravitational-wave detectors might be able to distinguish the type of the event and, in particular, identify the secondary, i.e., lighter component purely based on the GW signal~\citep{Chen:2020fzm, Fasano:2020eum}, this was not possible for GW190814 due to the large mass ratio $q\equiv m_1/m_2 \geq 1$ of the event\footnote{Please note that the definition of the mass ratio used here is the inverse of the definition used in \cite{Abbott:2020khf}.}. 
The tidal deformability of a binary black hole (BBH) system, $\tilde{\Lambda}=0$, is almost indistinguishable from an NS--BH merger with $\tilde{\Lambda}=\frac{16}{13} \Lambda_{\rm NS} (1+12q)/(1+q)^5 \lesssim 10^{-2}$ for the given system parameters, where $\Lambda_{\rm NS}$ is the NS tidal deformability.
In addition, the missing electromagnetic counterpart does not provide additional information because from an NS--BH system with such a heavy primary component no electromagnetic signal is likely to be detected, unless the BH has a very high spin, $\chi= c J/(G m^2)$ with speed of light $c$, angular momentum $J$, gravitational constant $G$ and the object's mass $m$ (e.g., \cite{Foucart:2012nc} and \cite{Kruger:2020gig}).
This is disfavored for GW190814 with the primary spin magnitude bounded to be $\chi_1<0.07$ at 90\% confidence~\citep{Abbott:2020khf}. 
Therefore, from observations alone, it cannot be determined if the secondary component of GW190814 is the lightest BH or the heaviest NS discovered to date, and its nature needs to be constrained differently.
Using the GW170817-informed EOS samples of \cite{Abbott:2018exr}, obtained with a spectral EOS parameterization~\citep{Lindblom:2010bb, Lindblom:2012zi}, the LVC found the probability for GW190814 to be an NS--BH merger is less than 3\%~\citep{Abbott:2020khf}. 
Using EOS-independent pulsar-mass distributions~\citep{Farr_2020}, they also reported a probability of less than 29\%.

Additional information on the nature of the secondary component of GW190814 might be obtained by considering the many new pieces of NS data obtained in the last years.
Besides mass measurements and observations of gravitational waves from BNS mergers, improved nuclear-physics constraints with uncertainty estimates from chiral effective field theory (EFT;~\cite{Hebeler:2013nza, Lynn:2016, Drischler:2017wtt}), recent X-ray observations by NICER~\citep{Miller:2019cac, Riley:2019yda}, or detailed modeling of the kilonova associated with GW170817~\citep{Kasen:2017sxr,Bulla:2019muo} allow us to reduce the uncertainty on the EOS~\citep{Annala:2017llu, Most:2018hfd, Greif:2018njt, Capano:2019eae, Raaijmakers:2019dks, Chatziioannou:2019yko, Dietrich:2020lps, Landry:2020vaw, Essick:2020flb}, see~\cite{Chatziioannou:2020pqz} for a recent review.
For example, \cite{Essick:2020ghc} addressed GW190814
by using a mass-based classification scheme employing Bayesian model selection and informed by compact-object populations and posteriors on $M_{\rm{max}}$ from the EOS model of \cite{Landry:2020vaw}, which includes information from mass measurements, BNS mergers, and NICER.
They found the probability of GW190814 to be an NS--BH merger to be less than 6\%, and less than 0.1\% when they additionally enforced the limit on $M_{\rm{max}}$ from \cite{Shibata:2019ctb}.

Here, we go further and use the Nuclear Physics -- Multimessenger Astrophysics (NMMA) framework developed in \cite{Dietrich:2020lps} to analyze GW190814 and identify the nature of its secondary component.
Our NMMA framework employs all of the additional sources of observational data (NS masses, GW data from GW170817 and GW190425, NICER data, and detailed kilonova modeling of GW170817) as well as EOS constrained by modern nuclear-physics theory, and hence, presents the first systematic multimessenger analysis of GW190814 using a wealth of interdisciplinary input.
We point out that the inclusion of multiple channels in our analysis provides the most complete understanding of GW190814, in contrast to previous studies that used only a subset of possible constraints.
Moreover, our analysis is based on a framework with controlled systematic uncertainties for all of its components~\citep{Dietrich:2020lps}.
We investigate the two different scenarios and also determine their constraints on the NS EOS. 
Our study strongly suggests that GW190814 was a BBH and not an NS--BH merger, see also similar conclusions in \cite{Abbott:2020khf}, \cite{Most:2020bba}, and \cite{Essick:2020ghc}.
To be conservative, we perform our analysis with and without assuming upper limits on $M_{\rm{max}}$ obtained from GW170817.
We find that the existence of a heavy NS in GW190814 leads to tension with current nuclear-physics constraints, see also \cite{Tan:2020ics} and \cite{1805759}.


\section{Analysis}

\begin{figure}[t]
\centering
\includegraphics[width=0.5\textwidth]{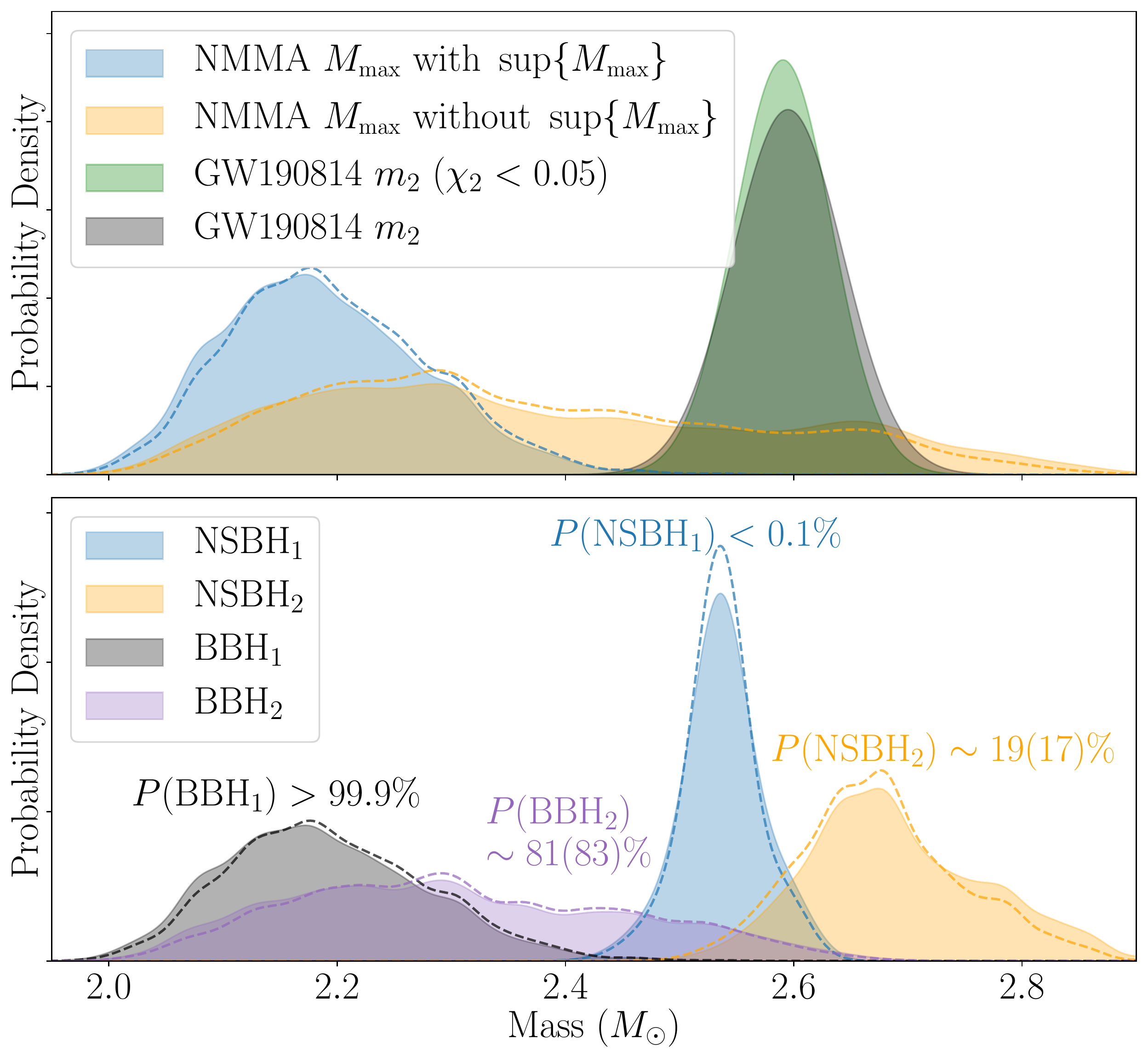}
\caption{Upper panel: posteriors on the maximum mass of NSs, $M_{\rm{max}}$, from the NMMA analysis of Ref.~\cite{Dietrich:2020lps} when enforcing the upper limit on $M_{\rm{max}}$ from Ref.~\cite{Rezzolla:2017aly} (blue) and when relaxing this constraint (orange). 
\red{We also show the posteriors for an EOS prior that is not flat in $R_{1.4}$ (dashed lines).}
We compare with the posterior on the mass of the secondary component of GW190814, assuming its spin $\chi_2<0.05$ (green) or with no limit on $\chi_2$ (black). 
Lower panel: the resulting posteriors on $M_{\rm{max}}$ for our four scenarios \red{with their probabilities. The probabilities for the alternative EOS prior are shown in brackets.}}
\label{fig:Mmaxposteriors}
\end{figure}

Our analysis starts from the NMMA framework introduced in \cite{Dietrich:2020lps}.
This approach is based on a set of 5000 EOSs that are constrained below $1.5$ times the nuclear saturation density, $n_{\rm{sat}}~\approx~2.7~\times~10^{14}$ g cm$^{-3}$ by state-of-the-art microscopic calculations using chiral EFT~\citep{Epelbaum:2008ga, Machleidt:2011zz}.
Chiral EFT is a systematic theory for nuclear interactions that allows us to quantify theoretical uncertainties~\citep{Epelbaum:2015epja, Drischler:2020yad}, and is valid at densities below $1-2 n_{\rm{sat}}$~\citep{Tews:2018kmu, Essick:2020flb}, although the exact breakdown of chiral EFT is an open problem.
Our NMMA EOS set is constrained below $1.5 n_{\rm{sat}}$ by quantum Monte Carlo (QMC) calculations~\citep{Carlson:2015} employing local chiral EFT interactions of \cite{Gezerlis:2014} and \cite{Lynn:2016} with systematic uncertainties.
Since NSs explore densities of several times $n_{\rm{sat}}$, we extend the EOSs beyond $1.5 n_{\rm{sat}}$ using the parametric speed-of-sound extension scheme developed in \cite{Tews:2018chv} and \cite{Tews:2019cap}, but see also \cite{Alford:2013aca} and \cite{Greif:2018njt} for different speed-of-sound schemes, \cite{Hebeler:2013nza} and \cite{RaithelOzel2016} for polytropic, \cite{Lindblom:2010bb} and \cite{Lindblom:2012zi} for spectral, and \cite{Landry:2018prl}, \cite{Essick:2020flb}, and \cite{Landry:2020vaw} for nonparametric extension schemes.
The chosen extension scheme is very versatile and includes exotic physics, e.g., strong first-order phase transitions.
Varying the density up to which microscopic constraints are enforced between $1-2 n_{\rm{sat}}$ does not significantly change the maximum mass posterior once astrophysical data is included~\citep{Essick:2020flb}.
Therefore, our choice of $1.5 n_{\rm{sat}}$ is robust.

Using Bayesian inference, these EOSs are analyzed with respect to their agreement with the posteriors on $M_{\rm{max}}$ from heavy-pulsar observations~\citep{Demorest:2010bx, Antoniadis:2013pzd, Cromartie:2019kug}, the upper limit on $M_{\rm{max}}$ ($M_{\rm{max}}^{\rm{up}}=2.16^{+0.17}_{-0.15} M_{\odot}$) from  \cite{Rezzolla:2017aly}, which is consistent with the limits inferred in \cite{Margalit:2017dij}, \cite{Ruiz:2017due}, \cite{Rezzolla:2017aly}, and \cite{Shibata:2019ctb}, the full mass-radius (MR) posterior from the NICER observation of PSR J0030+0451~\citep{Miller:2019cac, Riley:2019yda}, the GW observations of GW170817~\citep{TheLIGOScientific:2017qsa, Abbott:2018wiz} and GW190425~\citep{Abbott:2020uma}, and the kilonova observation AT2017gfo~\citep{GBM:2017lvd}, where both the pulsar mass measurements and the upper limit on $M_{\rm{max}}$ are approximated with Gaussian likelihoods.
Hence, the posterior of the NMMA analysis takes into account a wealth of available data on NSs; see~\cite{Dietrich:2020lps} for detailed information and discussions.
From the 5000 initial EOSs analyzed in \cite{Dietrich:2020lps}, about 20\% are within the 95\% credible interval given all observational constraints.
Our analyses allowed us to constrain the radius of a typical $1.4 M_{\odot}$ NS, $R_{1.4}$, to be $R_{1.4}=11.8^{+0.9}_{-0.8} \rm{ km}$~\citep{Dietrich:2020lps}.

\begin{figure}[t]
\centering
\includegraphics[width=0.5\textwidth]{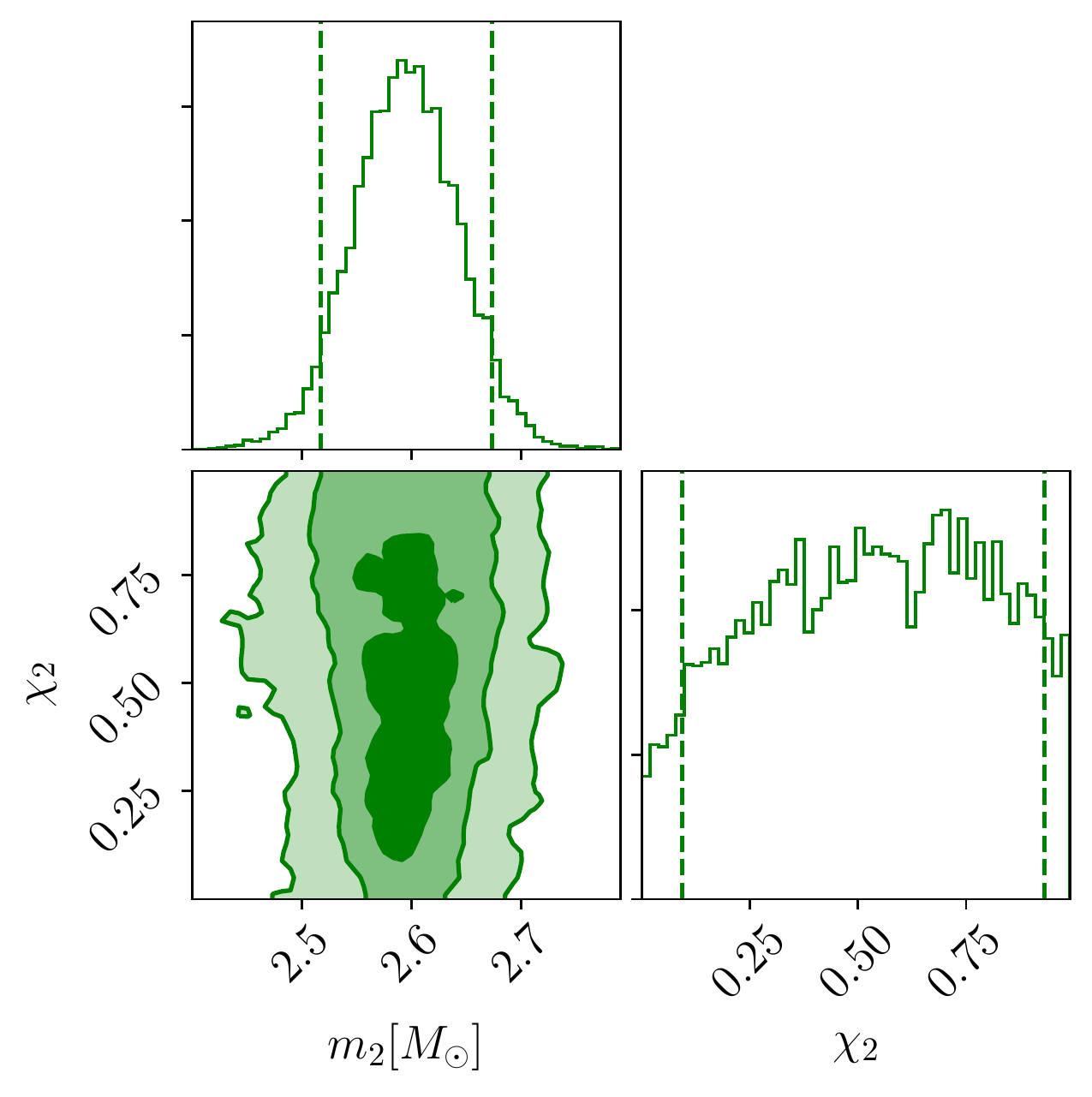}
\caption{Corner plot showing the posterior distribution of $m_2$--$\chi_2$ for GW190814. 
The dashed lines mark the 90\% credible intervals.}
\label{fig:m2a2posterior}
\end{figure}

We now use the final posterior of our NMMA analysis to investigate the nature of GW190814, which includes all analysis steps except the absence of EM emission from GW190425.
In particular, we study four scenarios.
In the first scenario, NSBH$_1$, we assume that GW190814 was an NS--BH merger.
Hence, GW190814 leads to an additional lower limit on $M_{\rm{max}}$.
Even though rotation could stabilize such a heavy NS against gravitational collapse for a lower $M_{\rm{max}}$~\citep{Rezzolla:2017aly, Most:2020bba, Zhang:2020, Tsokaros:2020hli}, the long lifetimes of BNS systems suggest NS spins to be small.
Therefore, we adopt the same low-spin prior used in the LVC studies, $\chi_2<0.05$~\citep{TheLIGOScientific:2017qsa}, suggesting $M_{\rm{max}}\geq 2.5 M_{\odot}$ at 90\% confidence.
Because upper limits on $M_{\rm{max}}$ from GW170817 might suffer from systematic uncertainties and are based on assumptions about the fate of the remnant, in the second scenario, NSBH$_2$, we again assume that GW190814 was an NS--BH merger but relax the upper bound on $M_{\rm{max}}$ of \cite{Rezzolla:2017aly}.
In our third scenario, BBH$_1$, we assume that GW190814 was a BBH merger, i.e., that the secondary component is a BH\footnote{It has been suggested that the secondary component might be a primordial black hole, see, e.g.,~\cite{Vattis:2020iuz}.}.
This scenario is the contrary to NSBH$_1$.
In this case, GW190814 leads to an additional upper limit on $M_{\rm{max}}$, consistent with the upper bound of \cite{Rezzolla:2017aly}. 
Finally, the scenario BBH$_2$ is the contrary to NSBH$_2$ and assumes GW190814 was a BBH merger but relaxes the $M_{\rm{max}}$ bound of \cite{Rezzolla:2017aly}.
The information on GW190814's secondary component's mass measurement is incorporated via the method described in \cite{Miller:2019nzo} with the posterior samples taken from \cite{GW190814_PE_samples}.


\section{Results} 

In the following, we discuss what we can learn about the nature of GW190814 and the dense-matter EOS probed in the core of NSs from our four scenarios.

\subsection{The nature of GW190814}

In Fig.~\ref{fig:Mmaxposteriors}, we compare the posterior on $M_{\rm{max}}$ of the NMMA analysis (blue), $M_{\rm max}=2.18^{+0.15}_{-0.13}M_{\odot}$, with the posterior of the mass of the secondary component of GW190814 depending on two choices for its spin: (1) $\chi_2<0.05$ (green), expected for NS, and (2) without any constraint on the spin (black), which is relevant in case of a BH.
Please note that the posterior widens slightly if higher spins are allowed, see Fig.~\ref{fig:m2a2posterior}.
Furthermore, we show the posterior of the NMMA analysis when we relax the upper bound on $M_{\rm{max}}$ of \cite{Rezzolla:2017aly} (orange), $M_{\rm max}=2.30^{+0.34}_{-0.25}M_{\odot}$. 

\begin{figure*}
\centering
\includegraphics[width=0.249\textwidth]{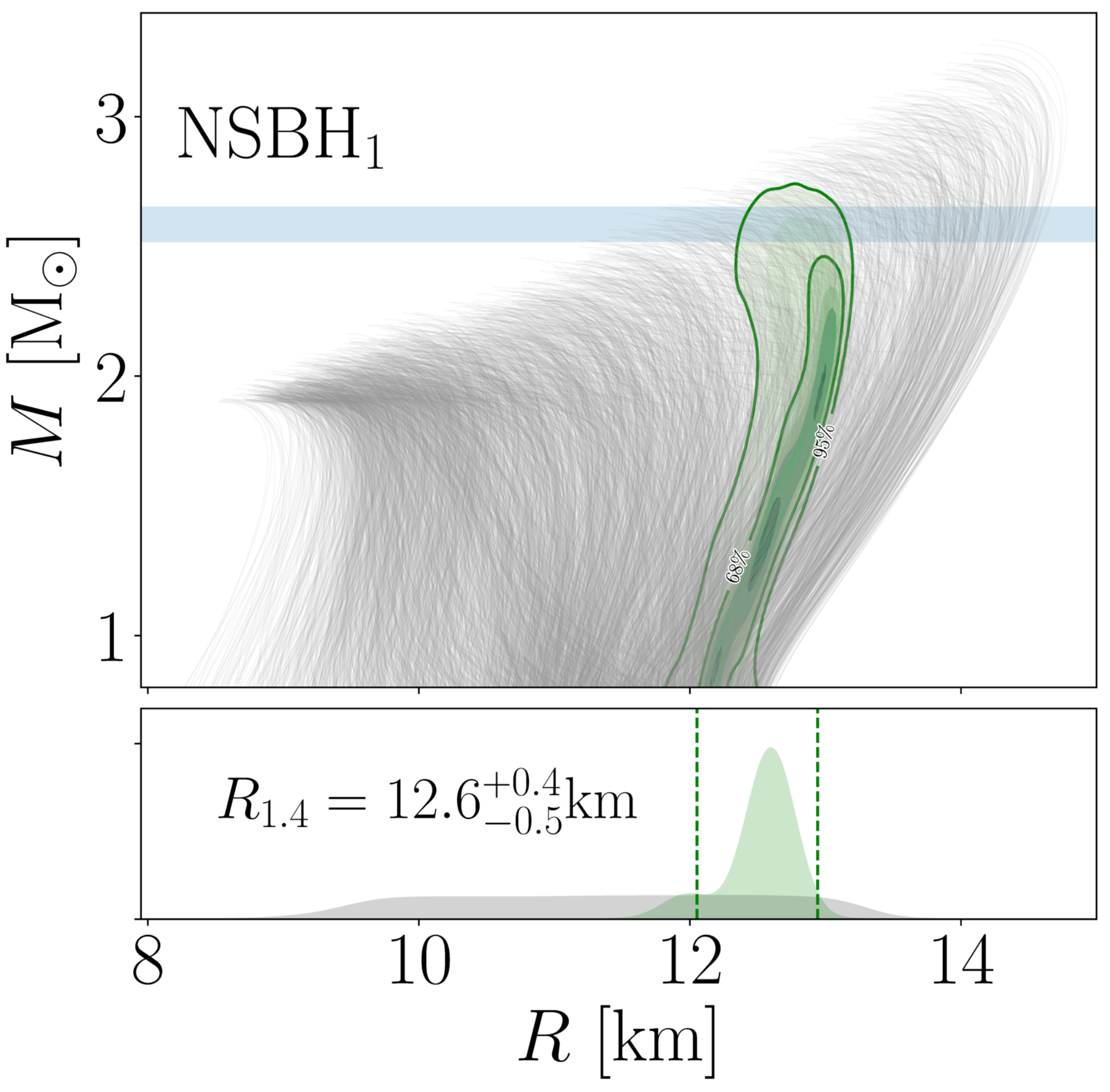}
\includegraphics[width=0.244\textwidth]{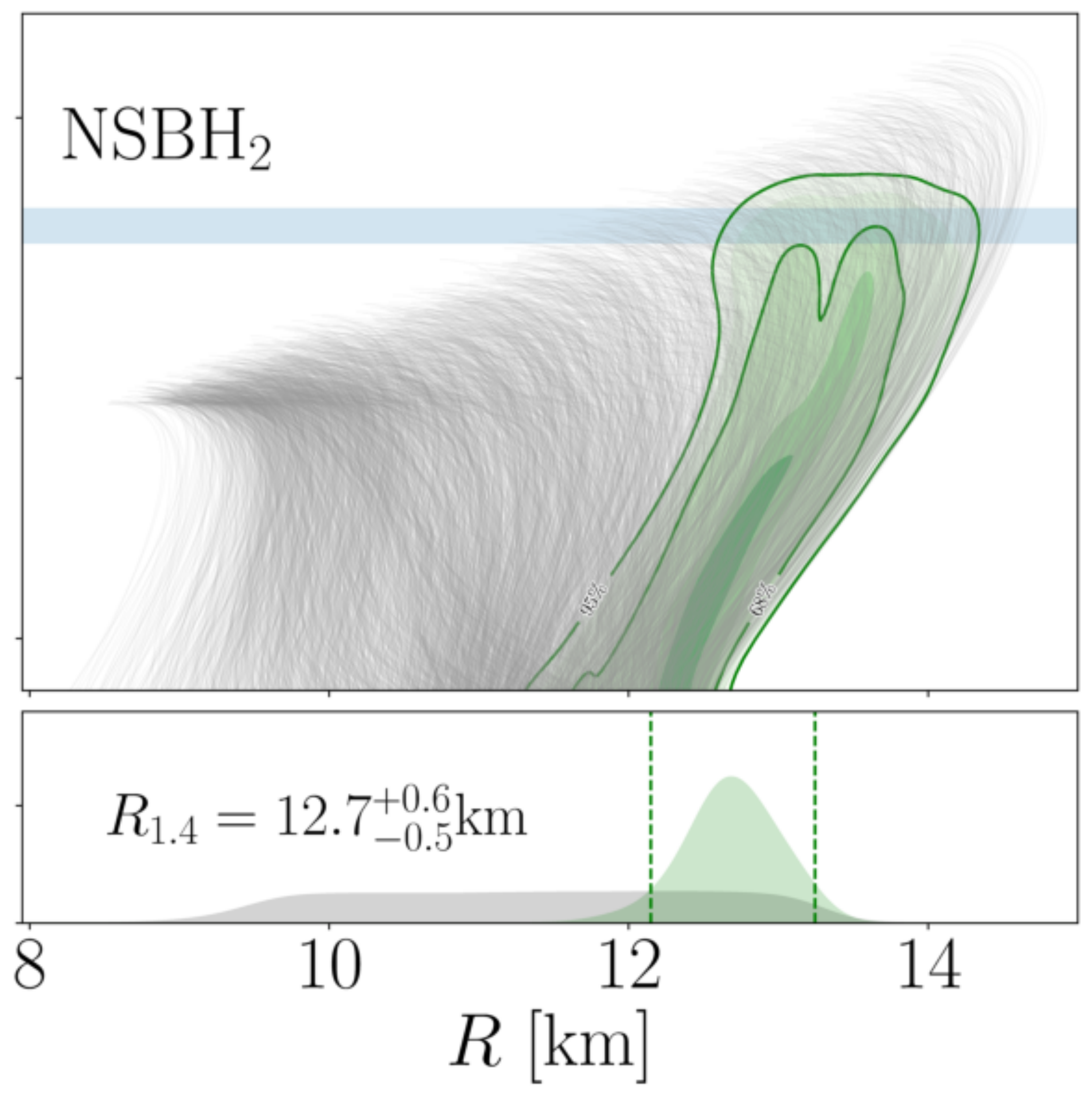}
\includegraphics[width=0.244\textwidth]{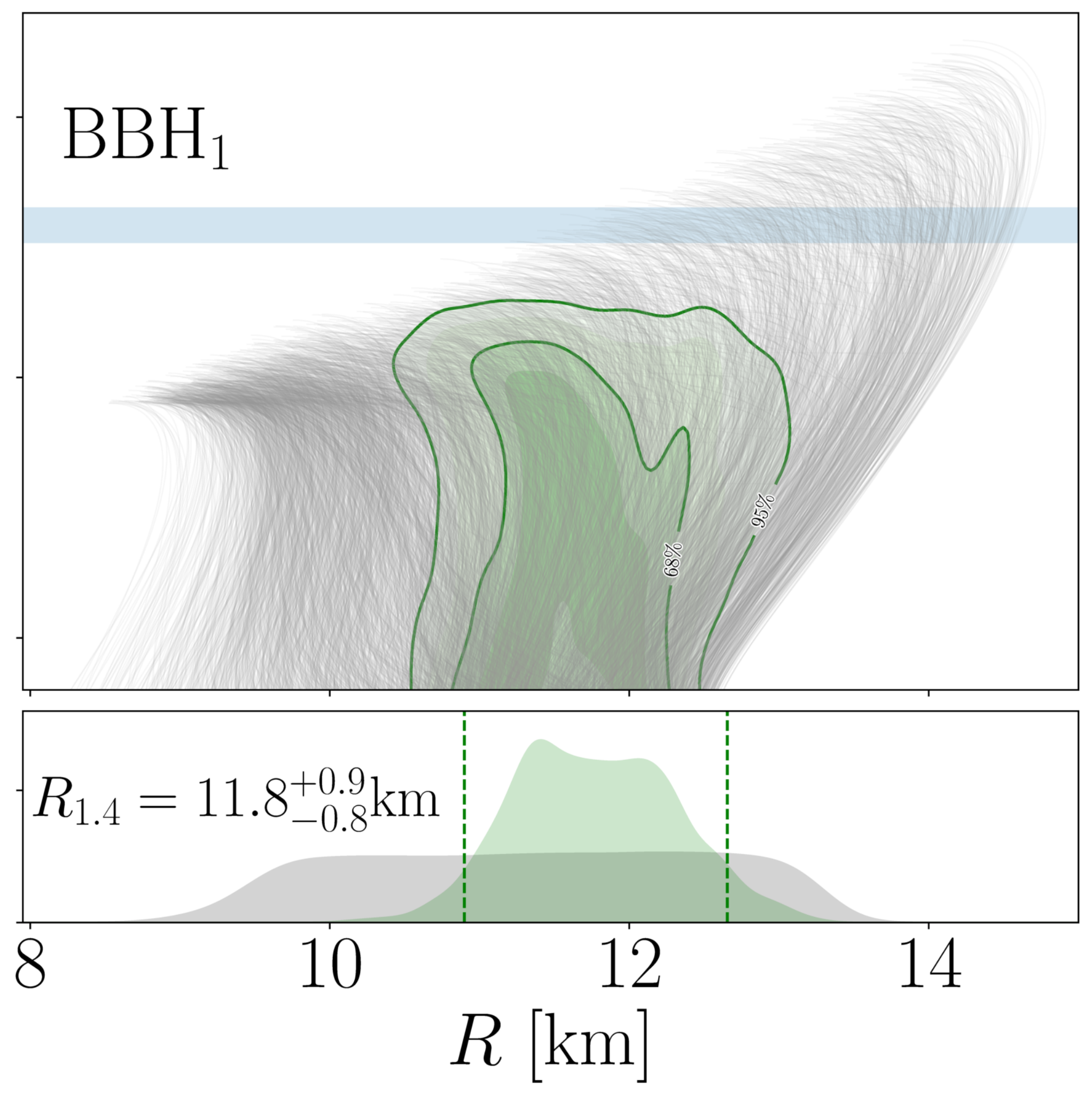}
\includegraphics[width=0.244\textwidth]{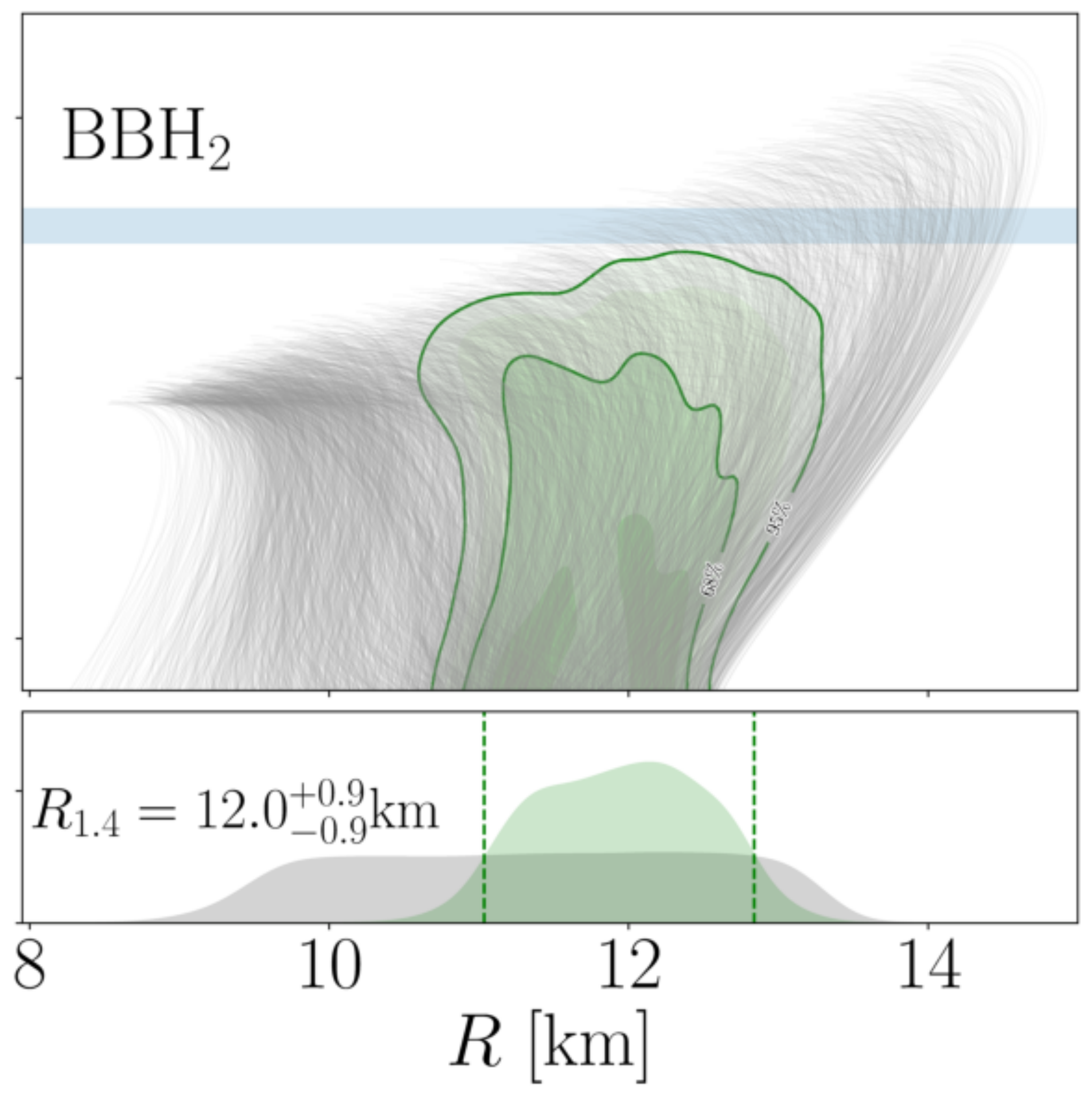}
\caption{Mass-radius relations for NSBH$_1$ (left), NSBH$_2$ (middle left), BBH$_1$ (middle right) and BBH$_2$ (right), and the corresponding constraint on the radius of a typical neutron star, $R_{1.4}$. 
We show all EOS in our set (gray) and the 95\% credible interval on the mass-radius relations that survive in each scenario (green-shaded areas). 
For comparison, we also show the 90\% credible interval of the mass of GW190814's secondary component (blue). 
For $\rm NSBH_1$ and $\rm NSBH_2$, a constraint of $\chi_2<0.05$ is imposed, while it is relaxed for $\rm BBH_1$ and $\rm BBH_2$.}
\label{fig:MRplot}
\end{figure*}

NSBH$_1$ is described by the overlap of the NMMA and mass posteriors (blue and green).
Due to the great tension between the upper limit on $M_{\rm{max}}$ extracted from the BNS merger GW170817~\citep{Rezzolla:2017aly} and the assumption that this new object, close to the remnant mass of GW170817, is an NS, this is the most restrictive of the four scenarios.
From the overlap of both posteriors, we can estimate the probability that the secondary component of GW190814 had a mass below $M_{\rm{max}}$ by using
\begin{equation}
    \begin{aligned}
    P(\Delta m >0) = \int_{0}^{\infty}d\Delta m \int_{-\infty}^{\infty} dm\, p_{M_{\rm{max}}}(m+\Delta m)p_{m_2}(m),
    \end{aligned}
\end{equation}
where $\Delta m \equiv M_{\rm{max}} - m_2$, and $p_{M_{\rm{max}}}(m)$ and $p_{m_2}(m)$ are the probability distribution for $M_{\rm{max}}$ from the NMMA analysis and the posterior on $m_2$ from GW190814, respectively.
The overlap region is extremely small, and hence, the probability $P(\Delta m >0)$ is less than $0.1\%$, in excellent agreement with \cite{Essick:2020ghc} with an upper limit on $M_{\rm{max}}$. 
Neutron-star EOS for this maximum-mass range are heavily penalized by the upper $M_{\rm{max}}$ limit.
In the NSBH$_1$ case, we find that the resulting $M_{\rm{max}}$ is constrained to a very narrow range, $M_{\rm{max}}=\NSBHoneMmax$; see the lower panel of Fig.~\ref{fig:Mmaxposteriors}.
This finding is in good agreement with the limit on $M_{\rm{max}}$ obtained from spin-polarized neutron matter in \cite{Tews:2019qhd}. 
However, this $M_{\rm{max}}$ would imply that the remnant of GW170817 was either a supramassive or long-lived NS, which is in conflict with the observed kilonova lightcurve and the GRB afterglow, e.g., \cite{Margalit:2017dij}.

For NSBH$_2$, the relaxation of the upper bound on $M_{\rm{max}}$ widens the NMMA posterior, see the orange curve in Fig.~\ref{fig:Mmaxposteriors}.
However, $M_{\rm{max}}$ is still constrained to be $M_{\rm{max}}\lesssim 2.8 M_{\odot}$ because of the required softness of the EOS at low densities to be consistent with GW170817, and the requirement for the EOS to explain the kilonova observations within our kilonova models, that depend on $M_{\rm{max}}$ through disk ejecta~\citep{Coughlin:2018fis, Dietrich:2020lps}.
While GW190814 again leads to an additional lower limit on $M_{\rm{max}}$, similarly to NSBH$_1$, the overlap region is now larger. 
This results in an increase of $P(\Delta m >0)$ to about $19\%$.
For NSBH$_2$, we find that $M_{\rm{max}}$ is constrained to $M_{\rm{max}}=\NSBHtwoMmax$.

NSBH$_2$ is also the ideal case to highlight the importance of the various aspects of our NMMA analysis. 
When including only nuclear-physics constraints and heavy-pulsar masses, $P(\Delta m >0)\sim 33\%$ which even increases to $\sim 44\%$ when NICER data are included.
This is because heavy pulsars favor high $M_{\rm{max}}$ and NICER data prefers stiff EOS that also tend to have higher $M_{\rm{max}}$.
The limitations on $M_{\rm{max}}$, and hence $P(\Delta m >0)$, stem from the inclusion of the GW signal GW170817 and its associated kilonova, which reduce $P(\Delta m >0)$ to $\sim 27\%$ and $\sim 20\%$, respectively. 
Finally, GW190425 reduces $P(\Delta m >0)$ to the $19\%$ quoted above.

For the scenario BBH$_1$, the observation of GW190814 adds an upper limit on $M_{\rm{max}}$, which now excludes the small overlap region of the blue and black curves in Fig.~\ref{fig:Mmaxposteriors} (note that for BBH$_{1,2}$ we do not constrain $\chi_2$). 
In this sense, BBH$_1$ is the contrary of NSBH$_1$, and the probability for the mass $m_2$ of GW190814 to be above $M_{\rm{max}}$, $P(\Delta m <0)$, is above 99.9\%.
As expected, this scenario does not visibly impact the posterior on $M_{\rm{max}}$ from the NMMA analysis and we obtain $M_{\rm{max}}=\BBHoneMmax$.
Finally, BBH$_2$ is the contrary to NSBH$_2$.
In this case, we find $P(\Delta m <0)\sim 81$\% and $M_{\rm{max}}=\BBHtwoMmax$.

We have also tested the dependence of our findings on the EOS prior. 
The NMMA analysis used an EOS prior that is uniform in $R_{1.4}$~\citep{Dietrich:2020lps}.
Using the EOS prior provided directly by the parametric speed-of-sound extension scheme developed in \cite{Tews:2018chv, Tews:2019cap}, i.e., an EOS prior that is nonuniform in $R_{1.4}$, we show the results as dashed lines in Fig.~\ref{fig:Mmaxposteriors}.
We find that the probability for NSBH$_2$ changes only very slightly for a different EOS prior, from 19\% to 17\%.
Hence, we conclude that our findings are robust with respect to the EOS prior.

We summarize the findings for all four scenarios in Table~\ref{tab:results}.
Our analysis strongly suggests that GW190814 was a BBH merger, as otherwise GW190814 would introduce a strong tension with the results of our NMMA analysis.
Given the remnant mass of GW170817 of $2.7 M_{\odot}$, which is very close to the inferred $m_2$ of GW190814, and the likely scenario that the remnant of GW170817 was, in fact, a BH, this seems to be the most consistent scenario given all current observational and theoretical knowledge of the NS EOS.
Our findings are consistent with \cite{Abbott:2020khf}, \cite{Tan:2020ics}, \cite{Most:2020bba}, and \cite{Essick:2020ghc}.

\subsection{EOS constraints from GW190814}

Finally, we investigate the impact of our four scenarios on the EOS by studying the MR relation.
This allows us to provide testable predictions for the NS radius for these four cases, and might help to fully pin down the nature of GW190814 when more observations become available in the future.

\begin{table*}
\centering
\tabcolsep=0.50cm
\def\arraystretch{1.9}
\caption{Summary of the probabilities and the resulting posteriors on the NS maximum mass and radius of a typical NS for the four scenarios analyzed in this work.}
\label{tab:results}
\begin{tabular}{c|cccc}
\hline
 Scenario & NSBH$_1$  & NSBH$_2$ & BBH$_1$ & BBH$_2$\\
\hline
Probability & $<0.001$ & $\sim 0.19$ & $>0.999$ & $\sim 0.81$ \\
$M_{\rm{max}}$ &  
$\NSBHoneMmax$ & $\NSBHtwoMmax$ &  $\BBHoneMmax$ & $\BBHtwoMmax$\\
$R_{1.4}$ & $\NSBHoneRonefour$ & $\NSBHtwoRonefour$ &  $\BBHoneRonefour$ & $\BBHtwoRonefour$\\
\hline
\end{tabular}
\end{table*}

For NSBH$_1$ ($P(\Delta m >0)<0.1$ \%), the strong tension between the different constraints on $M_{\rm{max}}$ translates to a narrow posterior in the MR plane, which we show in the left panel of Fig.~\ref{fig:MRplot}.
We find that in this case $R_{1.4}$ is constrained to be $R_{1.4}=\NSBHoneRonefourSim$. 
That would be the most stringent constraint on the NS radius to date.
Also, the posterior on MR remains rather tight in the mass range $(1.4-2.0) M_{\odot}$ and, hence, puts very strong constraints on the NS EOS.
For NSBH$_2$ ($P(\Delta m >0) \sim 19$ \%), the constraints on the MR relation are less tight and the posterior widens as expected, see middle left panel of Fig.~\ref{fig:MRplot}.
The radius of a typical NS is found to be $R_{1.4}=\NSBHtwoRonefourSim$, in good agreement with the determinations by the LVC~\citep{Abbott:2020khf} and \cite{Essick:2020ghc} for this scenario.
Both for the NSBH$_1$ and NSBH$_2$ scenarios, the EOS posterior strongly suggests that the NS MR relation does not have multiple stable branches, which would be indicative of very strong first-order phase transitions~\citep{Alford:2013aca}. 
For such EOS, $M_{\rm{max}}$ is typically much smaller, see, e.g., \cite{Alford:2013aca}, \cite{Alvarez-Castillo:2017qki}, and \cite{Chatziioannou:2019yko}. 
Furthermore, in particular NSBH$_2$ now prefers the stiffest EOS included in our EOS set, and its posterior is pushed against the upper bound of our EOS prior.
This behavior is observed also for the pressure between $(1-2) n_{\rm{sat}}$, which is pushed toward the upper prior bound.
Therefore, NSBH$_2$ might imply that the QMC calculations employing local chiral EFT interactions, which we use to constrain the EOSs, might break down already below $1.5n_{\rm{sat}}$.
In particular, NSBH$_2$ would disfavor the softer Hamiltonian explored in \cite{Tews:2018kmu, Tews:2018chv}, although it would not exclude it.
A possible explanation could be that higher-order many-body forces, that tend to stiffen the EOS~\citep{Tews:2012fj, Drischler:2020yad}, are crucial to describe NS physics, see also \cite{Essick:2020flb}.
 
Furthermore, NSBH$_2$ suggests that the EOS would need to remain stiff within the whole NS.
While the original NMMA analysis finds the maximum of the speed of sound inside an NS, $c_{s,\rm{max}}$, to be $c_{s,\rm{max}}^2\geq 0.4$, for NSBH$_2$ we find $c_{s,\rm{max}}^2\geq 0.6$.
Hence, GW190814 being an NS--BH merger might require us to revisit our current understanding of the EOS.

In the BBH$_1$ and BBH$_2$ scenarios, GW190814 adds another upper limit on $M_{\rm{max}}$, which however, is much weaker than the upper limit of \cite{Rezzolla:2017aly}.
Hence, for BBH$_1$, GW190814 does not add any additional information and our result of \cite{Dietrich:2020lps},  $R_{1.4}=\BBHoneRonefourSim$, is reproduced. 
For BBH$_2$, due to the limit on $M_{\rm{max}}$ being weaker, the radius posterior shifts to slightly larger radii, and we find $R_{1.4}=\BBHtwoRonefourSim$, which remains very consistent with the findings in \cite{Dietrich:2020lps}.
Because this new upper limit would be more robust than the one of \cite{Rezzolla:2017aly}, the BBH$_2$ scenario would provide a verification of the findings of the NMMA analysis if this scenario was confirmed.


\section{Summary}

In this Letter, we have investigated different possible scenarios for the nature of GW190814 using our robust NMMA framework that includes a wealth of observational data. 
Assuming that this compact merger was, in fact, an NS--BH merger, we find strong constraints on the radius of a typical neutron star, $R_{1.4}=\NSBHoneRonefourSim$ ($R_{1.4}=\NSBHtwoRonefourSim$) in case upper limits on $M_{\rm{max}}$ from GW170817 are (not) enforced.
If, on the other hand, GW190814 was a BBH merger, then it is fully consistent with our current knowledge of the EOS, and the radius of a typical NS remains at $R_{1.4}=\BBHoneRonefourSim$ ($R_{1.4}=\BBHtwoRonefourSim$). 

Based on the low probability of $m_2$ to lie below $M_{\rm{max}}$ inferred from the NMMA analysis of \cite{Dietrich:2020lps}, less than 0.1\% (19\%) if the upper limit on $M_{\rm{max}}$ of \cite{Rezzolla:2017aly} is (not) included, our study strongly suggests that GW190814 was a BBH merger.
Similar events detected in the future will help to map out the maximum mass of NSs and enable us to pin down the EOS of dense matter.

\acknowledgements
We thank N. Andersson, R. Essick, P. Landry, and J. Margueron for insightful discussions.
The work of I.T. was supported by the U.S. Department of Energy, Office of Science, Office of Nuclear Physics, under contract No.~DE-AC52-06NA25396, by the Laboratory Directed Research and Development program of Los Alamos National Laboratory under project number 20190617PRD1, and by the U.S. Department of Energy, Office of Science, Office of Advanced Scientific Computing Research, Scientific Discovery through Advanced Computing (SciDAC) program.
P.T.H.P is supported by the research program of the Netherlands Organization for Scientific Research (NWO). 
M.W.C. acknowledges support from the National Science Foundation with grant No. PHY-2010970.
S.A. is supported by the CNES Postdoctoral Fellowship at Laboratoire AstroParticule et Cosmologie.
J.H. acknowledges support from the National Science Foundation with grant No. PHY-1806990.
Computations have been performed on the Minerva HPC cluster of the Max-Planck-Institute for Gravitational Physics
and on SuperMUC-NG (LRZ) under project number pn56zo. 
Computational resources have also been provided by the Los Alamos National Laboratory Institutional Computing Program, which is supported by the U.S. Department of Energy National Nuclear Security Administration under Contract No.~89233218CNA000001, and by the National Energy Research Scientific Computing Center (NERSC), which is supported by the U.S. Department of Energy, Office of Science, under contract No.~DE-AC02-05CH11231.
This research has made use of data, software and/or web tools obtained from the Gravitational 
Wave Open Science Center (https://www.gw-openscience.org), a service of LIGO Laboratory, the 
LIGO Scientific Collaboration and the Virgo Collaboration. LIGO is funded by the U.S. National 
Science Foundation. Virgo is funded by the French Centre National de Recherche Scientifique (CNRS), 
the Italian Istituto Nazionale della Fisica Nucleare (INFN) and the Dutch Nikhef, with contributions 
by Polish and Hungarian institutes.

\bibliography{refs}

\begin{thebibliography}{}
\expandafter\ifx\csname natexlab\endcsname\relax\def\natexlab#1{#1}\fi
\providecommand{\url}[1]{\href{#1}{#1}}

\bibitem[{Aasi {et~al.}(2015)}]{TheLIGOScientific:2014jea}
Aasi, J., {et~al.} 2015, Class. Quant. Grav., 32, 074001

\bibitem[{Abbott {et~al.}(2017{\natexlab{a}})}]{GBM:2017lvd}
Abbott, B., {et~al.} 2017{\natexlab{a}}, Astrophys. J. Lett., 848, L12

\bibitem[{Abbott {et~al.}(2020{\natexlab{a}})}]{Abbott:2020uma}
---. 2020{\natexlab{a}}, Astrophys. J. Lett., 892, L3

\bibitem[{Abbott {et~al.}(2017{\natexlab{b}})}]{TheLIGOScientific:2017qsa}
Abbott, B.~P., {et~al.} 2017{\natexlab{b}}, Phys. Rev. Lett., 119, 161101

\bibitem[{Abbott {et~al.}(2018)}]{Abbott:2018exr}
---. 2018, Phys. Rev. Lett., 121, 161101

\bibitem[{Abbott {et~al.}(2019)}]{Abbott:2018wiz}
---. 2019, Phys. Rev., X9, 011001

\bibitem[{Abbott {et~al.}(2020{\natexlab{b}})}]{Abbott:2020khf}
Abbott, R., {et~al.} 2020{\natexlab{b}}, Astrophys. J., 896, L44

\bibitem[{Acernese {et~al.}(2015)}]{TheVirgo:2014hva}
Acernese, F., {et~al.} 2015, Class. Quant. Grav., 32, 024001

\bibitem[{Alford {et~al.}(2013)Alford, Han, \& Prakash}]{Alford:2013aca}
Alford, M.~G., Han, S., \& Prakash, M. 2013, Phys. Rev. D, 88, 083013

\bibitem[{Alsing {et~al.}(2018)Alsing, Silva, \& Berti}]{Alsing:2017bbc}
Alsing, J., Silva, H.~O., \& Berti, E. 2018, Mon. Not. Roy. Astron. Soc., 478,
  1377

\bibitem[{Alvarez-Castillo \& Blaschke(2017)}]{Alvarez-Castillo:2017qki}
Alvarez-Castillo, D., \& Blaschke, D. 2017, Phys. Rev. C, 96, 045809

\bibitem[{Annala {et~al.}(2018)Annala, Gorda, Kurkela, \&
  Vuorinen}]{Annala:2017llu}
Annala, E., Gorda, T., Kurkela, A., \& Vuorinen, A. 2018, Phys. Rev. Lett.,
  120, 172703

\bibitem[{Antoniadis {et~al.}(2013)Antoniadis, Freire, Wex, Tauris, Lynch,
  {et~al.}}]{Antoniadis:2013pzd}
Antoniadis, J., Freire, P.~C., Wex, N., {et~al.} 2013, Science, 340, 6131

\bibitem[{Arzoumanian {et~al.}(2018)}]{Arzoumanian:2017puf}
Arzoumanian, Z., {et~al.} 2018, Astrophys. J. Suppl., 235, 37

\bibitem[{Bulla(2019)}]{Bulla:2019muo}
Bulla, M. 2019, Mon. Not. Roy. Astron. Soc., 489, 5037

\bibitem[{Capano {et~al.}(2020)Capano, Tews, Brown, Margalit, De, Kumar, Brown,
  Krishnan, \& Reddy}]{Capano:2019eae}
Capano, C.~D., Tews, I., Brown, S.~M., {et~al.} 2020, Nature Astron., 4, 625

\bibitem[{Carlson {et~al.}(2015)Carlson, Gandolfi, Pederiva, Pieper,
  Schiavilla, Schmidt, \& Wiringa}]{Carlson:2015}
Carlson, J., Gandolfi, S., Pederiva, F., {et~al.} 2015, Rev. Mod. Phys., 87,
  1067

\bibitem[{Chatziioannou(2020)}]{Chatziioannou:2020pqz}
Chatziioannou, K. 2020, Gen. Rel. Grav., 52, 109

\bibitem[{Chatziioannou \& Han(2020)}]{Chatziioannou:2019yko}
Chatziioannou, K., \& Han, S. 2020, Phys. Rev. D, 101, 044019

\bibitem[{Chen {et~al.}(2020)Chen, Johnson-McDaniel, Dietrich, \&
  Dudi}]{Chen:2020fzm}
Chen, A., Johnson-McDaniel, N.~K., Dietrich, T., \& Dudi, R. 2020, Phys. Rev.
  D, 101, 103008

\bibitem[{Coughlin {et~al.}(2019)Coughlin, Dietrich, Margalit, \&
  Metzger}]{Coughlin:2018fis}
Coughlin, M.~W., Dietrich, T., Margalit, B., \& Metzger, B.~D. 2019, Monthly
  Notices of the Royal Astronomical Society: Letters, 489, L91

\bibitem[{Cromartie {et~al.}(2019)}]{Cromartie:2019kug}
Cromartie, H.~T., {et~al.} 2019, Nature Astron., 4, 72

\bibitem[{Demorest {et~al.}(2010)Demorest, Pennucci, Ransom, Roberts, \&
  Hessels}]{Demorest:2010bx}
Demorest, P., Pennucci, T., Ransom, S., Roberts, M., \& Hessels, J. 2010,
  Nature, 467, 1081

\bibitem[{Dietrich {et~al.}(2020{\natexlab{a}})Dietrich, Coughlin, Pang, Bulla,
  Heinzel, Issa, Tews, \& Antier}]{Dietrich:2020lps}
Dietrich, T., Coughlin, M.~W., Pang, P. T.~H., {et~al.} 2020{\natexlab{a}},
  Science, 370, 1450

\bibitem[{Dietrich {et~al.}(2020{\natexlab{b}})Dietrich, Hinderer, \&
  Samajdar}]{Dietrich:2020eud}
Dietrich, T., Hinderer, T., \& Samajdar, A. 2020{\natexlab{b}},
  arXiv:2004.02527

\bibitem[{Drischler {et~al.}(2019)Drischler, Hebeler, \&
  Schwenk}]{Drischler:2017wtt}
Drischler, C., Hebeler, K., \& Schwenk, A. 2019, Phys. Rev. Lett., 122, 042501

\bibitem[{Drischler {et~al.}(2020)Drischler, Melendez, Furnstahl, \&
  Phillips}]{Drischler:2020yad}
Drischler, C., Melendez, J.~A., Furnstahl, R.~J., \& Phillips, D.~R. 2020,
  Phys. Rev. C, 102, 054315

\bibitem[{Epelbaum {et~al.}(2009)Epelbaum, Hammer, \&
  Meissner}]{Epelbaum:2008ga}
Epelbaum, E., Hammer, H.-W., \& Meissner, U.-G. 2009, Rev. Mod. Phys., 81, 1773

\bibitem[{Epelbaum {et~al.}(2015)Epelbaum, Krebs, \&
  Mei\ss{}ner}]{Epelbaum:2015epja}
Epelbaum, E., Krebs, H., \& Mei\ss{}ner, U.-G. 2015, Eur. Phys. J. A, 51, 53.
\newblock \url{http://dx.doi.org/10.1140/epja/i2015-15053-8}

\bibitem[{Essick \& Landry(2020)}]{Essick:2020ghc}
Essick, R., \& Landry, P. 2020, Astrophys. J., 904, 80

\bibitem[{Essick {et~al.}(2020)Essick, Tews, Landry, Reddy, \&
  Holz}]{Essick:2020flb}
Essick, R., Tews, I., Landry, P., Reddy, S., \& Holz, D.~E. 2020, Phys. Rev. C,
  102, 055803

\bibitem[{Farr \& Chatziioannou(2020)}]{Farr_2020}
Farr, W.~M., \& Chatziioannou, K. 2020, Research Notes of the {AAS}, 4, 65.
\newblock \url{https://doi.org/10.3847%2F2515-5172%2Fab9088}

\bibitem[{Fasano {et~al.}(2020)Fasano, Wong, Maselli, Berti, Ferrari, \&
  Sathyaprakash}]{Fasano:2020eum}
Fasano, M., Wong, K. W.~K., Maselli, A., {et~al.} 2020, Phys. Rev. D, 102,
  023025

\bibitem[{Fattoyev {et~al.}(2020)Fattoyev, Horowitz, Piekarewicz, \&
  Reed}]{1805759}
Fattoyev, F.~J., Horowitz, C.~J., Piekarewicz, J., \& Reed, B. 2020, Phys. Rev.
  C, 102, 065805

\bibitem[{Fishbach {et~al.}(2020)Fishbach, Essick, \& Holz}]{Fishbach:2020ryj}
Fishbach, M., Essick, R., \& Holz, D.~E. 2020, Astrophys. J. Lett., 899, L8

\bibitem[{Foucart(2012)}]{Foucart:2012nc}
Foucart, F. 2012, Phys. Rev. D, 86, 124007

\bibitem[{Gezerlis {et~al.}(2014)Gezerlis, Tews, Epelbaum, Freunek, Gandolfi,
  Hebeler, Nogga, \& Schwenk}]{Gezerlis:2014}
Gezerlis, A., Tews, I., Epelbaum, E., {et~al.} 2014, Phys. Rev. C, 90, 054323

\bibitem[{Gill {et~al.}(2019)Gill, Nathanail, \& Rezzolla}]{Gill:2019bvq}
Gill, R., Nathanail, A., \& Rezzolla, L. 2019, Astrophys. J., 876, 139

\bibitem[{Greif {et~al.}(2019)Greif, Raaijmakers, Hebeler, Schwenk, \&
  Watts}]{Greif:2018njt}
Greif, S., Raaijmakers, G., Hebeler, K., Schwenk, A., \& Watts, A. 2019, Mon.
  Not. Roy. Astron. Soc., 485, 5363

\bibitem[{Hebeler {et~al.}(2013)Hebeler, Lattimer, Pethick, \&
  Schwenk}]{Hebeler:2013nza}
Hebeler, K., Lattimer, J., Pethick, C., \& Schwenk, A. 2013, Astrophys. J.,
  773, 11

\bibitem[{Kalogera \& Baym(1996)}]{Kalogera:1996ci}
Kalogera, V., \& Baym, G. 1996, Astrophys. J. Lett., 470, L61

\bibitem[{Kasen {et~al.}(2017)Kasen, Metzger, Barnes, Quataert, \&
  Ramirez-Ruiz}]{Kasen:2017sxr}
Kasen, D., Metzger, B., Barnes, J., Quataert, E., \& Ramirez-Ruiz, E. 2017,
  Nature, 10.1038/nature24453, arXiv:1710.05463

\bibitem[{Krüger \& Foucart(2020)}]{Kruger:2020gig}
Krüger, C.~J., \& Foucart, F. 2020, Phys. Rev. D, 101, 103002

\bibitem[{Landry \& Essick(2019)}]{Landry:2018prl}
Landry, P., \& Essick, R. 2019, Phys. Rev. D, 99, 084049

\bibitem[{Landry {et~al.}(2020)Landry, Essick, \&
  Chatziioannou}]{Landry:2020vaw}
Landry, P., Essick, R., \& Chatziioannou, K. 2020, Phys. Rev. D, 101, 123007.
\newblock \url{https://link.aps.org/doi/10.1103/PhysRevD.101.123007}

\bibitem[{Lattimer(2012)}]{Lattimer:2012nd}
Lattimer, J.~M. 2012, Ann. Rev. Nucl. Part. Sci., 62, 485

\bibitem[{Lindblom(2010)}]{Lindblom:2010bb}
Lindblom, L. 2010, Phys. Rev., D82, 103011

\bibitem[{Lindblom \& Indik(2012)}]{Lindblom:2012zi}
Lindblom, L., \& Indik, N.~M. 2012, Phys. Rev. D, 86, 084003

\bibitem[{Lynn {et~al.}(2016)Lynn, Tews, Carlson, Gandolfi, Gezerlis, Schmidt,
  \& Schwenk}]{Lynn:2016}
Lynn, J.~E., Tews, I., Carlson, J., {et~al.} 2016, Phys. Rev. Lett., 116,
  062501

\bibitem[{Machleidt \& Entem(2011)}]{Machleidt:2011zz}
Machleidt, R., \& Entem, D.~R. 2011, Phys. Rept., 503, 1

\bibitem[{Margalit \& Metzger(2017)}]{Margalit:2017dij}
Margalit, B., \& Metzger, B.~D. 2017, Astrophys. J., 850, L19

\bibitem[{Miller {et~al.}(2019{\natexlab{a}})Miller, Chirenti, \&
  Lamb}]{Miller:2019nzo}
Miller, M.~C., Chirenti, C., \& Lamb, F.~K. 2019{\natexlab{a}}, Astrophys. J.,
  888, 12.
\newblock \url{https://doi.org/10.3847/1538-4357/ab4ef9}

\bibitem[{Miller {et~al.}(2019{\natexlab{b}})}]{Miller:2019cac}
Miller, M.~C., {et~al.} 2019{\natexlab{b}}, Astrophys. J. Lett., 887, L24

\bibitem[{Most {et~al.}(2020)Most, Papenfort, Weih, \& Rezzolla}]{Most:2020bba}
Most, E.~R., Papenfort, L.~J., Weih, L.~R., \& Rezzolla, L. 2020, Mon. Not.
  Roy. Astron. Soc., 499, L82

\bibitem[{Most {et~al.}(2018)Most, Weih, Rezzolla, \&
  Schaffner-Bielich}]{Most:2018hfd}
Most, E.~R., Weih, L.~R., Rezzolla, L., \& Schaffner-Bielich, J. 2018, Phys.
  Rev. Lett., 120, 261103

\bibitem[{Raaijmakers {et~al.}(2020)}]{Raaijmakers:2019dks}
Raaijmakers, G., {et~al.} 2020, Astrophys. J. Lett., 893, L21

\bibitem[{{Raithel} {et~al.}(2016){Raithel}, {{\"O}zel}, \&
  {Psaltis}}]{RaithelOzel2016}
{Raithel}, C.~A., {{\"O}zel}, F., \& {Psaltis}, D. 2016, \apj, 831, 44

\bibitem[{Rezzolla {et~al.}(2018)Rezzolla, Most, \& Weih}]{Rezzolla:2017aly}
Rezzolla, L., Most, E.~R., \& Weih, L.~R. 2018, Astrophys. J., 852, L25

\bibitem[{Riley {et~al.}(2019)}]{Riley:2019yda}
Riley, T.~E., {et~al.} 2019, Astrophys. J. Lett., 887, L21

\bibitem[{Ruiz {et~al.}(2018)Ruiz, Shapiro, \& Tsokaros}]{Ruiz:2017due}
Ruiz, M., Shapiro, S.~L., \& Tsokaros, A. 2018, Phys. Rev., D97, 021501

\bibitem[{Shibata {et~al.}(2019)Shibata, Zhou, Kiuchi, \&
  Fujibayashi}]{Shibata:2019ctb}
Shibata, M., Zhou, E., Kiuchi, K., \& Fujibayashi, S. 2019, Phys. Rev., D100,
  023015

\bibitem[{Tan {et~al.}(2020)Tan, Noronha-Hostler, \& Yunes}]{Tan:2020ics}
Tan, H., Noronha-Hostler, J., \& Yunes, N. 2020, Phys. Rev. Lett., 125, 261104

\bibitem[{Tews {et~al.}(2018{\natexlab{a}})Tews, Carlson, Gandolfi, \&
  Reddy}]{Tews:2018kmu}
Tews, I., Carlson, J., Gandolfi, S., \& Reddy, S. 2018{\natexlab{a}},
  Astrophys. J., 860, 149

\bibitem[{Tews {et~al.}(2013)Tews, Krüger, Hebeler, \& Schwenk}]{Tews:2012fj}
Tews, I., Krüger, T., Hebeler, K., \& Schwenk, A. 2013, Phys. Rev. Lett., 110,
  032504

\bibitem[{Tews {et~al.}(2018{\natexlab{b}})Tews, Margueron, \&
  Reddy}]{Tews:2018chv}
Tews, I., Margueron, J., \& Reddy, S. 2018{\natexlab{b}}, Phys. Rev., C98,
  045804

\bibitem[{Tews {et~al.}(2019)Tews, Margueron, \& Reddy}]{Tews:2019cap}
---. 2019, Eur. Phys. J., A55, 97

\bibitem[{Tews \& Schwenk(2020)}]{Tews:2019qhd}
Tews, I., \& Schwenk, A. 2020, Astrophys. J., 892, 14

\bibitem[{{The LIGO Scientific Collaboration and the Virgo
  Collaboration}(2020)}]{GW190814_PE_samples}
{The LIGO Scientific Collaboration and the Virgo Collaboration}. 2020.
\newblock \url{https://dcc.ligo.org/P2000183/public}

\bibitem[{Tsokaros {et~al.}(2020)Tsokaros, Ruiz, \& Shapiro}]{Tsokaros:2020hli}
Tsokaros, A., Ruiz, M., \& Shapiro, S.~L. 2020, Astrophys. J., 905, 48

\bibitem[{Vattis {et~al.}(2020)Vattis, Goldstein, \&
  Koushiappas}]{Vattis:2020iuz}
Vattis, K., Goldstein, I.~S., \& Koushiappas, S.~M. 2020, Phys. Rev. D, 102,
  061301

\bibitem[{Zhang \& Li(2020)}]{Zhang:2020}
Zhang, N.-B., \& Li, B.-A. 2020, Astrophys. J., 902, 38

\end{thebibliography}

\end{document}